# Effects of Grain Boundaries and Defects on Anisotropic Magnon Transport in Textured $Sr_{14}Cu_{24}O_{41}$

Xi Chen,[1] Karalee Jarvis,[1] Sean Sullivan,[1] Yutao Li,[1] Jianshi Zhou,[1,2] and Li Shi[1,2,*]

[1] Materials Science and Engineering Program, Texas Materials Institute, The University of Texas at Austin, Austin, Texas 78712, USA
[2] Department of Mechanical Engineering, The University of Texas at Austin, Austin, Texas 78712, USA
*lishi@mail.utexas.edu

**Abstract:** The strong spin-spin exchange interaction in some low-dimensional magnetic materials can give rise to a high group velocity and thermal conductivity contribution from magnons. One example is the incommensurate layered compounds $(Sr,Ca,La)_{14}Cu_{24}O_{41}$. The effects of grain boundaries and defects on quasi-one-dimensional magnon transport in these compounds are not well understood. Here we report the microstructures and anisotropic thermal transport properties of textured $Sr_{14}Cu_{24}O_{41}$, which are prepared by solid-state reaction followed by spark plasma sintering. Transmission electron microscopy clearly reveals nano-layered grains and the presence of dislocations and planar defects. The thermal conductivity contribution and mean free paths of magnons in the textured samples are evaluated with the use of a kinetic model for one-dimensional magnon transport, and found to be suppressed significantly as compared to single crystals at low temperatures. The experimental results can be explained by a one-dimensional magnon-defect scattering model, provided that the magnon-grain boundary scattering mean free path in the anisotropic magnetic structure is smaller than the average length of these nano-layers along the *c* axis. The finding suggests low transmission coefficients for magnons across grain boundaries.

## I. Introduction

Materials with high thermal conductivity ($\kappa$) can be used to address the thermal management challenge in microelectronics and other functional devices and systems.[1] High $\kappa$ values can be found in some metals due to a large electronic contribution, which is proportional to the electrical conductivity according to the Wiedemann-Franz law. In comparison, record high thermal conductivity values have been found in different carbon allotropes including diamond, graphite, graphene, and carbon nanotubes, where the light elements and strong covalent bonding result in a large thermal conductivity contribution from the lattice vibrations or phonons. In addition, recent first-principles theoretical calculations have predicted that cubic boron arsenide can possess an even higher $\kappa$ value than diamond at high temperatures due to its unique phonon dispersion,[2] which has motivated experimental investigations.[3, 4]

Recently, in some low-dimensional magnetic materials with strong exchange interactions, $\kappa$ has been found to reach a reasonably high magnitude as a result of the collective spin excitations, the energy quantum of which is called a magnon.[5-7] Several cuprates with high magnon thermal conductivity ($\kappa_{mag}$) have been investigated, such as spin chain compounds $SrCuO_2$ and $Sr_2CuO_3$,[8] spin ladder compounds $(Sr,Ca,La)_{14}Cu_{24}O_{41}$,[9, 10] pseudo-two-ladder compound $CaCu_2O_3$,[11] and two-dimensional antiferromagnet $La_2CuO_4$.[12] In previous studies of $Sr_{14}Cu_{24}O_{41}$ by inelastic neutron scattering (INS), the antiferromagnetic coupling energy for the ladder sublattice along the ladder direction was found to be considerably larger than that along the rung.[14] Steady-state thermal conductivity measurements of $Sr_{14}Cu_{24}O_{41}$ single crystals reveal a large anisotropic ratio in $\kappa$. The thermal conductivity along the $c$-axis ($\kappa_c$) exhibits a second broad peak at 140 K with a maximum value of ~100 $Wm^{-1}K^{-1}$, which can be attributed to the magnetic excitations propagating along the spin ladders.[9, 10] In comparison, the $\kappa$ along the $a$ or $b$ axis show the typical behavior of the phonon contribution, which is only about 3-7 $Wm^{-1}K^{-1}$ at 140 K.

Prior studies on magnon heat transport in spin ladder compounds have been limited to single crystals. Based on a kinetic model, the measured $\kappa$ of $Sr_{14}Cu_{24}O_{41}$ single crystals was used to estimate the magnon mean free math (MFP), which was found to be about several hundred nanometers at temperatures below 150 K.[10] It was further demonstrated that magnon-hole scattering plays an important role in the magnon transport in these spin ladder compounds. For instance, the $\kappa_{mag}$ is suppressed significantly in the hole-doped $Sr_{14-x}Ca_xCu_{24}O_{41,}$ [15] whereas the substitution of Sr with La reduces the hole concentration, leading to the increase of $\kappa_{mag}$.[10] Since growth of large single crystals of $(Sr,Ca,La)_{14}Cu_{24}O_{41}$ is challenging due to its incongruent melting behavior, large-size polycrystalline materials of these compounds are potentially needed for applications in either thermal management or emerging spin caloritronic devices.[16, 17] However, the effects of grain boundaries and defects on the magnon thermal conductivity in polycrystalline structures are not well understood. In addition, although magnon scattering by defects has been investigated in several prior theoretical works,[18-22] these theories remain to be validated by experiments.

In this Article, we report a study of the effects of grain boundaries and defects on magnon thermal transport in relatively large-size textured $Sr_{14}Cu_{24}O_{41}$ samples. These textured samples are synthesized by solid-state reaction (SSR) followed by spark plasma sintering (SPS). X-ray diffraction (XRD) and electron microscopy studies reveal that the nanosized layered structures, which are grown along the *ac* plane, are preferably aligned with the *c* axis normal to the SPS press direction, leading to an anisotropic $\kappa$. The magnon MFPs of the polycrystalline samples are evaluated using a kinetic model of one-dimensional (1D) magnon thermal transport in conjunction with a theory of 1D magnon-defect scattering. The theoretical model can explain the experimental results provided that the magnon-grain boundary scattering length is smaller than the average grain size along the *c* axis as a result of small transmission coefficients of magnons at grain boundaries.

## II. Experimental Methods

Powders of $SrCO_3$ (99.95%) and CuO (99.99%) purchased from Alfa Aesar were used as the starting materials. Mixture of $SrCO_3$ and CuO with Sr:Cu ratio of 14:24 was heated at 900 $^{o}$C for 40 h in air. Subsequently, the obtained $Sr_{14}Cu_{24}O_{41}$ power was loaded into a graphite die with an inner diameter of 12 mm and consolidated into dense pellets by SPS under 50 MPa, resulting in a cylinder with a diameter of 12 mm and a thickness of ~18 mm. The as-pressed sample was then loaded into a graphite die with a larger inner diameter of 20 mm and then repressed by SPS under 50 MPa in vacuum. The pellet with a diameter of 20 mm and a thickness of ~4 mm was obtained. The texture is formed as a result of lateral material flow perpendicular to the press direction.[23] In this study, we have prepared two textured samples at different temperatures during SPS. Sample 1 is prepared by SPS at 750 $^{o}$C for 10 min and subsequently repressed at 780 $^{o}$C for 10 min. Sample 2 is fabricated by SPS at 750 $^{o}$C for 10 min and repressed at 850 $^{o}$C for 10 min. In comparison, a polycrystalline sample with nearly random grain orientation (sample 3) is prepared by one-step SPS at 800 $^{o}$C for 10 min. Finally, the pellets were annealed in air at 900 $^{o}$C for 40 h in order to improve purity.

The phase and crystal structure of the samples were studied by XRD with a Philips X'pert diffractometer with Cu K$\alpha$ radiation ($\lambda$=1.54184 Å). To study the texture effect, the XRD measurements were performed on the repressed pellets in the planes along and perpendicular to the press direction, respectively. The pole figure measurement was carried on a Rigaku Ultima IV diffractometer. The Archimedes method was used to measure the density of the samples as 5.01 g cm$^{-3}$ for sample 1, 5.28 g cm$^{-3}$ for sample 2, and 4.60 g cm$^{-3}$ for sample 3. The microstructures of the samples were analyzed by using a Quanta 600 environmental scanning electron microscopy (SEM) and a JEOL 2010F transmission electron microscopy (TEM). TEM specimens were prepared by cutting a 3 mm disc using ultrasonic disk cutter (Fischione, model 170) and polished to 200 $\mu$m in thickness. The center was polished using a dimple grinder down to a 0.06 μm Silica finish. The final preparation was conducted using a Gatan precision ion polishing system. The thermal conductivity of the samples was measured in the temperature interval between ~10 K and 300 K by a steady-state method.[24] To study the thermal conductivity of the samples in both directions parallel and perpendicular to the SPS press direction, the bar-shaped samples of about 0.5 $\times$ 0.5 $\times$ 3 mm were cut from the repressed pellets from both directions. The reference was a rod of constantan alloy with a diameter of 0.5 mm.

The differential thermocouple was made of copper and constantan wires. The uncertainty of the thermal conductivity measurement is about 15%.

## III. Results and Discussions

### A. Crystal structure and microstructure of textured $Sr_{14}Cu_{24}O_{41}$

The crystal structure of $Sr_{14}Cu_{24}O_{41}$ consists of two incommensurate sublattices,[25] as shown in Figure 1a. One is the chain sublattice with nearly 90° Cu-O-Cu bonds. The other is the two-leg ladder sublattice, where Cu forms nearly 180° Cu-O-Cu bonds along the *a* and *c* axes and each ladder is coupled by 90° Cu-O-Cu bonds. Planes of the chains are stacked alternately with planes of the ladders along *b* and are separated by strings of the Sr ions. The lattice parameters are $a$ = 11.469 Å, $b$ = 13.368 Å, $c_L$=3.931 Å for the ladders along the *c* axis and $c_C$ =2.749 Å for the chains along the *c* axis.[13] The Cu $S$=1/2 spins in $Cu_2O_3$ ladders form strong antiferromagnetic coupling via the 180° Cu-O-Cu superexchange, while the character of the superexchange intercation in $CuO_2$ $S$ = 1/2 spin chains is changed into a much weaker ferromagnetic coupling due to the 90° Cu-O-Cu configuration.[13]

Figure 1b shows the XRD patterns of sample 1 after SSR, after SPS, and after annealing, respectively. A relatively pure $Sr_{14}Cu_{24}O_{41}$ phase was formed after SSR. However, some $SrCuO_2$ and CuO phases were observed after SPS. Because the SPS was performed in vacuum and the sample was covered by the graphite paper or die, a certain amount of $Sr_{14}Cu_{24}O_{41}$ has been reduced as $Sr_{14}Cu_{24}O_{41} \rightarrow 14SrCuO_2 + 10CuO + (3/2)O_2\uparrow$. After annealing the pellets at 900 °C in air following the SPS step, a relatively pure $Sr_{14}Cu_{24}O_{41}$ phase was formed again, as indicated by the XRD results in Figure 1b.

To investigate the grain orientation, XRD analysis was performed on the planes both perpendicular (⊥) and parallel (//) to the press direction, as shown in Figure 1c. The diffraction intensity of the (002) peak measured on the plane perpendicular to the press direction is much stronger than that on the plane parallel to the press direction. This result indicates that the reorientation of (00*l*) planes of the grains into the pellet plane took place during the repressing process. We have further calculated the orientation factor $F$ of (00*l*) diffractions according to

$$F = \frac{p - p_0}{1 - p_0} , \tag{1}$$

$$p = \frac{\sum I(00l)}{\sum I(hkl)} , \tag{2}$$

$$p_0 = \frac{\sum I_0(00l)}{\sum I_0(hkl)} , \tag{3}$$

where $I(hkl)$ and $I_0(hkl)$ are the peak intensities for the SPS samples in this work and the calculated diffraction pattern of a randomly oriented sample, respectively.[26] The calculated $F$ values are 0.12 for sample 1 and 0.21 for sample 2, respectively. It is clear that the higher temperature during SPS can enhance the texture formation. Highly textured samples may be achieved by further increasing the temperature during SPS, which is still kept below the decomposition temperature of $Sr_{14}Cu_{24}O_{41}$, which is about 970 $^{o}$C.

Pole figures of the pellets after SPS were measured to achieve a better understanding of the texture information. As shown in Figure 2a-c, a maximum intensity appears at the center of the (002) pole figure when measured across the press direction while not in the data measured along the press direction, suggesting that the (00*l*) plane is preferentially aligned perpendicular to the press direction. Figure 2d-f show the intensity along the line cutting across the center of the pole figures. Sample 2 shows a stronger intensity in the center, which is due to the annealing effect at a higher temperature during SPS. The pole figure measurement clearly demonstrates that the samples after SPS are textured with a *c*-axis preferred orientation perpendicular to the press direction.

Figure 3 shows the SEM images of the fractured surface of sample 2 after SPS. As shown in Figure 3a and b, the fractured surface contains particle features with a lateral size of about 3 μm. The high-resolution SEM images in Figure 3c and d further reveal the presence of nanosized layers in each particle. These layers should be related to the weak interaction of incommensurate ladder and chain sublattices along the *b* axis. It has also been found that single crystal of $Sr_{14}Cu_{24}O_{41}$ can be cleaved easily along the *ac* plane.[27]

TEM study has been carried out to characterize the microstructures of the SPS samples. Figure 4a shows the typical TEM image of the layered microstructure. The width of each layer ranges from 20 to 100 nm. Figure 4b displays the bright-field TEM of these nano-layers, which suggests that defects are formed, as highlighted by red cycles. Further investigations reveal two types of defects, planar defects and dislocations (Figure 4c-4f). The selected area electron diffraction (SAED) pattern in Figure 4c shows spot splitting, which is common for a low angle grain boundary between two layers. This observed spot splitting indicates that these two nano-layers have a very small angle of misorientation (~1 degree) along the *b* axis. These layers are found to be grown along the *ac* plane and aligned along the *b* axis, as confirmed by SAED and high-resolution TEM (HRTEM). The typical grain size of these layers along the *c* axis is over 200 nm, as shown in the representative TEM results of Figure 4a, 5a, and Figure S1 in the Supporting Information. Furthermore, dislocation pairs form at the grain boundaries, as clearly seen in the filtered image of the (020) planes (Figure 4e). The several dislocations observed in the TEM measurements show an extra half-plane of atoms parallel to the *c* axis, as revealed in Figure 4e and Figure S2 in the Supporting Information. Twin boundaries are also formed as shown in Figure 4f. HRTEM shows two parallel incoherence twin boundaries with a spacing of ~30 nm. The angle between the matrix and the twin boundary is about 74 degrees.

Figure 5a shows the highly distorted nano-layers and highly-strained grain boundaries between two nanosized layers in the SPS sample. As shown in Figure 5b, the reflections associated with the ladder sublattice shows a small angle (~3 degrees) with respect to those from the chain sublattice, indicating that these two sublattices are not perfectly aligned along the *b* axis. Such orientation anomaly has also been observed in incommensurate higher manganese silicides.[28] The ladder sublattices can show a rotation with respect to the chain sublattices within the *ac* plane as seen in the SAED. The zone axis of the ladder sublattice is [105], while the zone axis of the chain sublattice is [100]. These defects observed in the sintered pellets can be formed due to the mass diffusion during solid-state reaction, disorder-order transitions, or mechanical deformation during SPS. These nano-layers grown along the *ac* plane, coupled with grain boundaries, twins and dislocations, can affect the magnon thermal transport in polycrystalline samples, as discussed below.

**B. Thermal conductivity of textured $Sr_{14}Cu_{24}O_{41}$**

Figure 6a shows the $\kappa$ of the two textured $Sr_{14}Cu_{24}O_{41}$ samples (samples 1 and 2) measured parallel and perpendicular to the press direction. As a result of texturing, the $\kappa$ values of polycrystalline samples are anisotropic. The measured $\kappa$ perpendicular to the press direction ($\kappa_{\perp}$) is considerably higher than that parallel to the press direction ($\kappa_{//}$). It should be noted that the $\kappa$ in the plane normal to the press direction is nearly isotropic as shown in Figure S3 (Supporting Information). The measured values of sample 2 are about 7.2 W $m^{-1}K^{-1}$ for $\kappa_{\perp}$ and 3.8 W $m^{-1}K^{-1}$ for $\kappa_{//}$ at 300 K, and 14.7 W $m^{-1}K^{-1}$ for $\kappa_{\perp}$ and 6.9 W $m^{-1}K^{-1}$ for $\kappa_{//}$ at 140 K. In addition, we have further annealed the same sample in oxygen at 900 $^{o}$C for 40 h after the initial thermal conductivity measurement. The $\kappa$ values measured after this second annealing step remain nearly the same as the initial results, as shown in Figure S3 (Supporting Information). This finding indicates that the oxygen stoichiometry in our samples has approached thermodynamic equilibrium after the first annealing step.

In order to better understand the effect of structural anisotropy on thermal transport, we have studied the thermal anisotropy ($\kappa_{\perp}/\kappa_{//}$) of $Sr_{14}Cu_{24}O_{41}$ samples. Figure 6b shows $\kappa_{\perp}/\kappa_{//}$ as a function of the orientation factor $F$ at 300 K. For samples with completely random orientation, the $F$ value is 0 and $\kappa_{\perp}/\kappa_{//}$ should be 1. When the $F$ value of $Sr_{14}Cu_{24}O_{41}$ single crystal is 1, $\kappa_{\perp}/\kappa_{//}$ can be calculated as $0.5(\kappa_a+\kappa_c)/\kappa_b$, which is about

4. Together with the present results, the $\kappa_{\perp}/\kappa_{//}$ increases with orientation factor. As shown in the inset of Figure 6b, the values for $\kappa_{\perp}/\kappa_{//}$ range from 1.6 to 2.2 in the measured temperature range. Sample 2 shows a larger $\kappa$ anisotropy due to a higher temperature during SPS.

### C. Magnon thermal transport analysis

We have analyzed the magnon thermal transport in these textured samples. The solid thermal conductivity ($\kappa_S$) of the textured $Sr_{14}Cu_{24}O_{41}$ sample with a small porosity ($\Phi$) can be calculated for each of the two transport directions as[29]

$$\kappa_{S,\perp,//} = \kappa_{\perp,//}\frac{2+\Phi}{2-2\Phi}\,. \qquad (4)$$

The porosity is 6.7%, 1.7%, and 14.3% for sample 1, 2, and, 3, respectively. The $\kappa_S$ of the three samples were corrected for porosity and shown in Figure 6c. The $\kappa_S$ below 40 K is dominated by the phonon contribution ($\kappa_L$) due to large magnon gaps and high electrical resistivity.[14, 30] We fit the data between 10 and 40 K with $\kappa_L$~1/(a$T$+b),[31] which is due to increasing umklapp process of phonons with increasing temperature. The magnon thermal conductivity is evaluated by subtracting $\kappa_L$ from the total $\kappa_S$, as displayed in Figure 6d for each of the two directions. The obtained $\kappa_{mag,\perp}$ perpendicular to the press direction is much higher than the obtained $\kappa_{mag,//}$ parallel to the press direction due to the preferred orientation of grains. At 150 K, the $\kappa_{mag,\perp}$ values are 7.5 $Wm^{-1}K^{-1}$ for sample 1 and 8.8 $Wm^{-1}K^{-1}$ for sample 2, while the $\kappa_{mag,//}$ are 3.0 $Wm^{-1}K^{-1}$ for sample 1 and 2.3 $Wm^{-1}K^{-1}$ for sample 2. Both $\kappa_{mag,\perp}$ and $\kappa_{mag,//}$ of the texture samples are much smaller than the reported $\kappa_{mag}$ for $Sr_{14}Cu_{24}O_{41}$ single crystal along the *c* axis.[10] For the textured samples, the intrinsic magnon thermal conductivity along the *c* axis can be calculated as $\kappa_{mag} = 2\kappa_{mag,\perp} + \kappa_{mag,//}$. [32] The obtained $\kappa_{mag}$ values are also considerably lower than the single crystal value, as shown in Figure 7a. For temperatures below ~100 K, $\kappa_{mag}$ first increases rapidly due to excitation of magnon above the spin gap, and decreases with further increase of temperature as a result of enhanced magnon scattering by other quasiparticles or defects. Therefore, a peak in $\kappa_{mag}$ is shown at about 150 K.

In order to gain further insight into the reduced $\kappa_{mag}$ in the textured samples, we have evaluated the average magnon mean free path (MFP), $l_{mag}$, from a kinetic model for 1D magnon transport[5, 10]

$$\kappa = \frac{1}{2\pi}\int c_k v_k l_k dk, \tag{5}$$

where the integration is over the allowable 1D wave vector ($k$) space, $c_k = \frac{dn_k}{dT}\varepsilon_k$ is the mode specific heat; $\varepsilon_k$ is the energy quantum, $n_k$ is the occupation function, $v_k$ is the velocity, and $l_k$ is the MFP, respectively, for the mode $k$. In addition, $v_k = (d\varepsilon/dk)/\hbar$, where $\hbar$ is the reduced Planck constant. Under the assumption of a frequency-independent MFP for the gapped one-dimensional two-leg ladder, Eq. (5) yields[5, 10]

$$\kappa_{mag} = \frac{3Nl_{mag}}{\pi\hbar k_B T^2}\int_{\Delta}^{\varepsilon_{max}} \frac{\exp(\frac{\varepsilon}{k_B T})}{\left[\exp(\frac{\varepsilon}{k_B T})+3\right]^2}\varepsilon^2 d\varepsilon, \tag{6}$$

where $N$ is the number of ladders per unit cross section area perpendicular to the ladders, $\Delta$ is the energy gap of the singlet-triplet dispersion of ladders, $\varepsilon$ is the energy, $\varepsilon_{max}$ is the band maximum of the dispersion, $T$ is the temperature, and the factor of three in the expression reflects the presence of spin triplets in the spin ladder. It should be noted that a particular form of magnon dispersion is not needed for deriving the thermal conductivity expression of a 1D magnon system. In our analysis, the values for $\Delta$ and $\varepsilon_{max}$ are 32.5 meV and 200 meV, respectively, according to a previous INS study.[14] Figure 7b compares the magnon MFP values calculated based on Eq. (6) for a single crystal and the polycrystalline samples in this work. The $l_{mag}$ of the single crystal along the *c* axis is very large, about 2200 Å at 100 K. With the increase of temperature, the $l_{mag}$ is reduced and approaches a constant value of ~60 Å at 300 K. For the polycrystalline samples, the $l_{mag}$ along the *c* axis is significantly suppressed especially at low temperatures, which are about 500 Å at 100 K. However, the $l_{mag}$ values of polycrystalline samples are comparable to that of single crystal at room temperature. These results reveal that grain boundary and defect scattering processes play a more important role at a lower temperature, whereas intrinsic magnon-magnon and magnon-phonon scattering are dominant at room temperature and above.

Next, we examine the origin for the much reduced $l_{mag}$ in the polycrystalline samples. To account for the grain size and defect effects on the magnon thermal conductivity, we include a boundary scattering MFP ($l_b$) and a defect scattering term ($l_d$) in the total MFP as

$$l_{mag}^{-1} = l_{single}^{-1} + l_b^{-1} + l_d^{-1}, \quad (7)$$

where $l_{single}$ is the magnon MFP of single crystal along the *c* axis. Without additional defect scattering, the $\kappa_{mag}$ is reduced with decreasing grain size as shown in Figure 7a. In addition, the peak of $\kappa_{mag}$ is shifted to slightly higher temperature. However, the shape of the calculated $\kappa_{mag}$ does not agree with $\kappa_{mag}$ without including defect scattering. According to Callaway *et al.*,[18, 19] $l_d^{-1}$ can be calculated as $l_d^{-1} = N_d A k^4$ for a three-dimensional (3D) isotropic system where $N_d$ is the number of magnetic defects, and $A$ is a constant related to the material's properties. This expression is similar to the phonon-defect scattering rate in a 3D system,[33] which also shows a $k^4$ dependence. The origin of this relationship is that the scattering rate depends on the wave vector as $k^2 D(k)$, [33-35] where $D(k)$ is the density of states and is proportion to $k^2$ for a 3D system. Because $D(k)$ is constant in 1D, the magnon-defect scattering rate should be proportional to $k^2$. The singlet-triplet dispersion relationship is approximated as $E(k) = \Delta + \hbar v k$, where $v$ is the group velocity of magnons, as such dispersion is quite linear accord to the INS studies.[14, 36, 37] Therefore, the magnon-defect scattering MFP should follow $l_d^{-1} = C_d\, k^2$ for 1D magnon transport in spin ladder compounds, where $C_d$ is a constant. By including both boundary and defect scattering, the agreement between the calculated $\kappa_{mag}$ and the experimental data is improved considerably, as shown in Figure 7a. The obtained optimum fitting parameters are $l_b$ = 108 nm and $C_d = 5.4 \times 10^{-11}$ m for sample 1, and $l_b$ = 118 nm and $C_d = 4.59 \times 10^{-11}$ m for sample 2, respectively. Varying $l_b$ by just 25% from the optimum values are found to result in poor fitting of the experimental data, as revealed by the sensitivity analysis in Figure S4 (Supporting Information).

Although the theoretical model of 1D magnon-defect scattering can help to explain the experimental data, the obtained $l_b$ is considerably smaller than the observed grain size along the *c* axis, which should be considerably larger than 200 nm based on the representative TEM images of Figure 4a, 5a and Figure S1 (Supporting Information).

TEM results have not shown grain boundaries or dislocations within 200 nm distances between them. In the case of phonon transport, it has been suggested that the phonon boundary scattering length ($l_{b,ph}$) is related to the average grain size ($D_{avg}$) as $l_{b,ph} = \alpha D_{avg}$, where $\alpha$ is the parameter accounting for the effect of grain boundary transmission of phonons.[38] It has been found that the value of $\alpha$ is less than one in some silicon,[38] bismuth telluride[29] and silicide[39] bulk samples because of low transmission coefficients of phonons across weakly bonded grain boundary. If a similar model is adopted for magnons, the parameter $\alpha$ for magnons would be only about 0.54-0.59 for the textured samples. Magnons can couple with magnons across the grain boundary both directly or via coupling with phonons at the same side of the boundary in conjunction with phonon transmission across grain boundaries. A grain boundary can lead to significant reduction of the short-range antiferromagnetic exchange interaction due to the distorted or broken bonds at the boundaries as well as the orientation fluctuation of the exchange interaction.[40] Therefore, direct magnon-magnon coupling can be as weak as phonon-phonon transmission across a weakly bounded grain boundary. In addition, unlike three-dimensional phonon transport, quasi-one-dimensional transport of magnons can result in very low direct transmission of magnons across the grain boundary of two randomly oriented grains. Moreover, the grain boundary scattering based on magnon-phonon coupling mechanism can be associated with significant thermal resistance across the magnon-phonon relaxation length, which can be as long as the order of 100 nm due to weak magnon-phonon coupling in the spin ladder compounds.[7] As a result of the low transmission coefficients for both magnons and phonons at grain boundaries as well as the large thermal resistance associated with weak magnon-phonon coupling, the obtained magnon-boundary scattering length can be smaller than the grain size along the *c* axis.

## IV. Conclusions

Relatively large-size textured $Sr_{14}Cu_{24}O_{41}$ polycrystalline samples have been obtained by pressing the samples using SPS for multiple times. The thermal conductivity perpendicular to the SPS press direction is found to be higher than that along the press direction. Such anisotropy of thermal conductivity is caused by the reorientation of the *c* axis of the small layered grains into the pellet plane. Electron microscopy studies reveal that $Sr_{14}Cu_{24}O_{41}$ nano-layers with width of 20-100 nm are grown along the *ac* plane and stacked along the *b* axis. A number of defects, including dislocations, planar defects, and the grains with orientation anomaly, are present in these textured samples. Moreover, the magnon MFPs of the textured samples have been evaluated from the measured thermal conductivity, and found to be suppressed considerably at low temperatures. The suppression can be explained by a defect-scattering model of 1D magnons, provided that the magnon-grain boundary scattering MFP is considerably shorter than the grain size along the spin ladder. The result points to very low transmission coefficients of one-dimensional magnons across grain boundaries for both direct magnon-magnon transmission and magnon-phonon-phonon-magnon transmission.

## Acknowledgements

The authors thank Steve Swinnea for the assistance of pole figure measurement, C. Hess for sharing the raw data of magnon thermal conductivity of the $Sr_{14}Cu_{24}O_{41}$ single crystal, and David G. Cahill and Annie Weathers for helpful discussions. This work was supported by US Army Research Office (ARO) MURI award W911NF-14-1-0016. The SPS processing at The University of Texas at Austin was conducted with instrument acquired with the support of NSF award number DMR-1229131.

## Figures

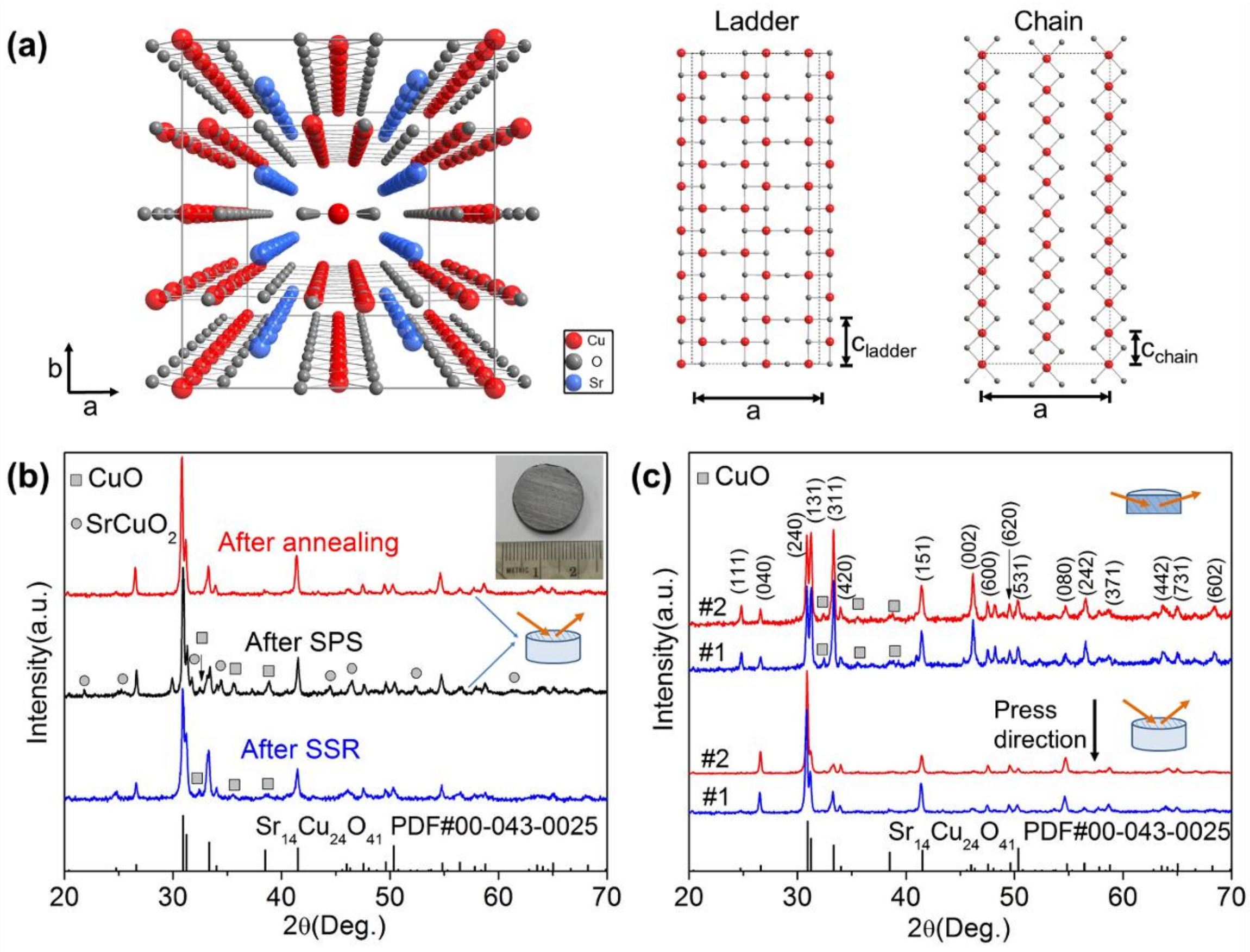

**Figure 1.** (Color online) (a) Crystal structure of the spin ladder compound $Sr_{14}Cu_{24}O_{41}$. (b) XRD patterns of the $Sr_{14}Cu_{24}O_{41}$ sample 1 after SSR, SPS and annealing. The inset of (b) is a photograph of a 20 mm diameter, 4 mm thick disc made by SPS. (c) XRD patterns of two $Sr_{14}Cu_{24}O_{41}$ samples taken on the planes parallel and perpendicular to the press direction.

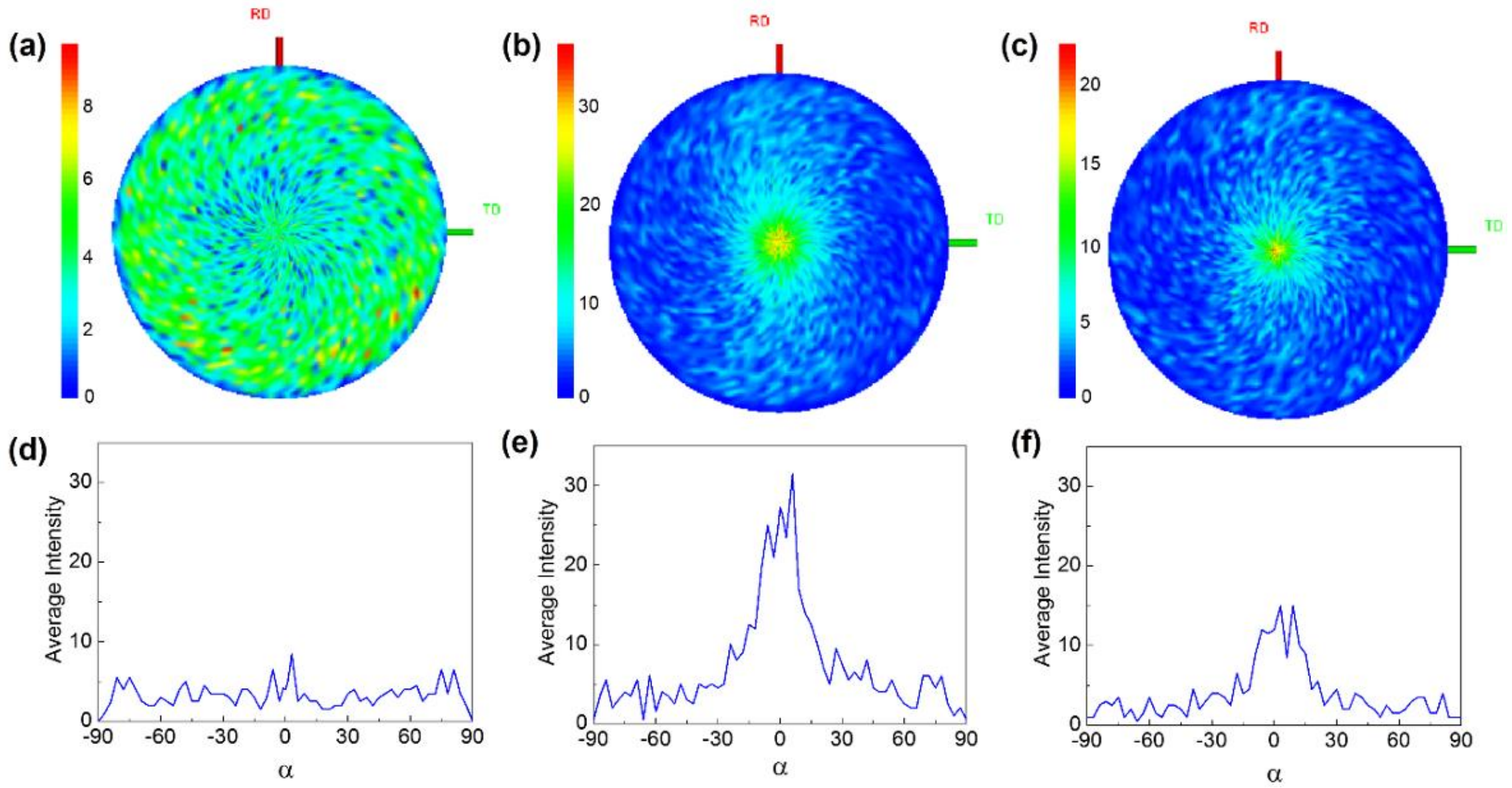

**Figure 2.** (Color online) (a,b) The (002) pole figures of sample 2 after SPS along and across the press direction, respectively. (c) The (002) pole figure of sample 1 after SPS across the press direction. (d-f) The corresponding line cuts (cross-sections of the pole figures) for (a), (b) and (c), respectively.

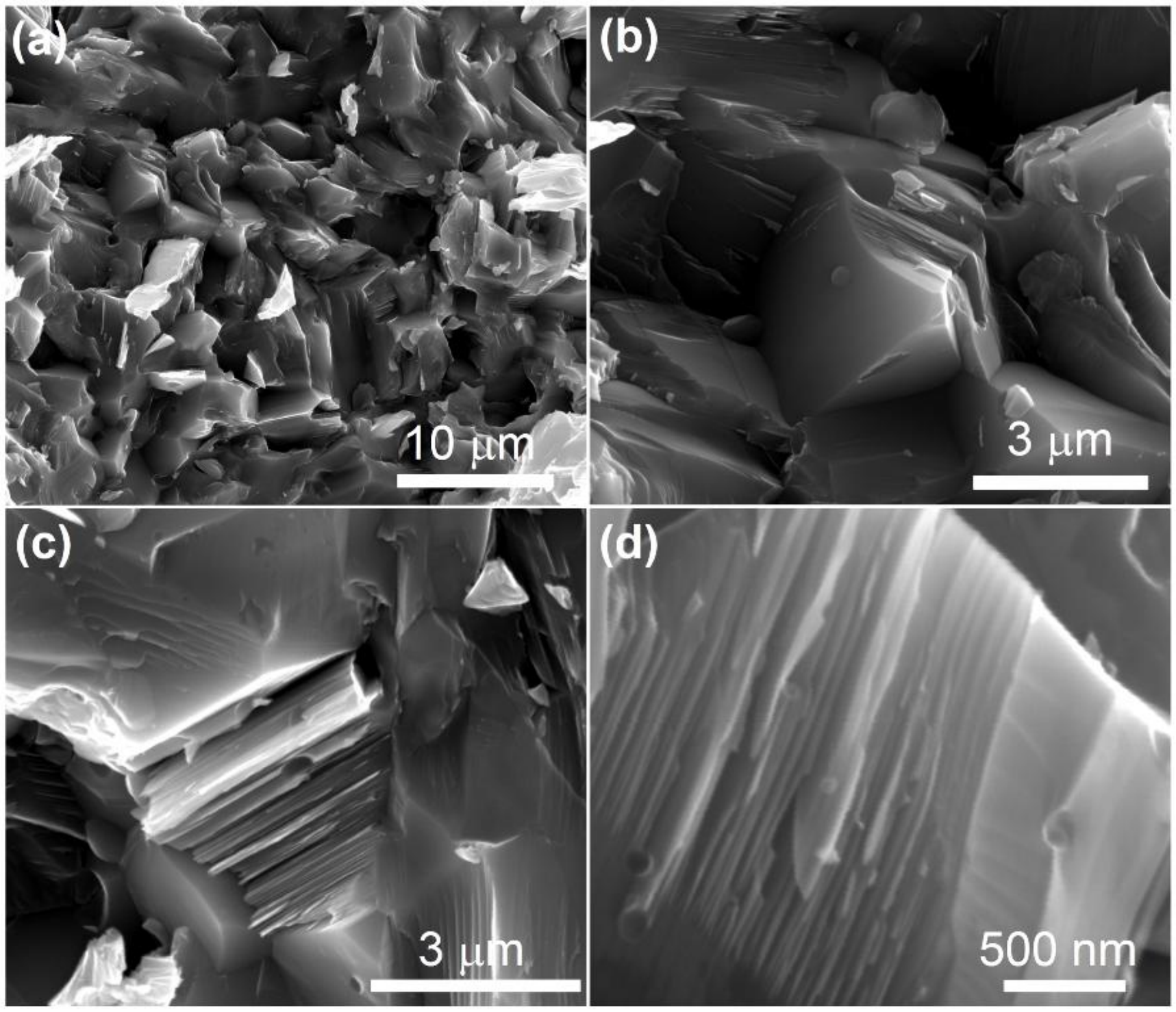

**Figure 3.** (Color online) (a-d) SEM images of the fracture surface of the $Sr_{14}Cu_{24}O_{41}$ sample 2 after annealing.

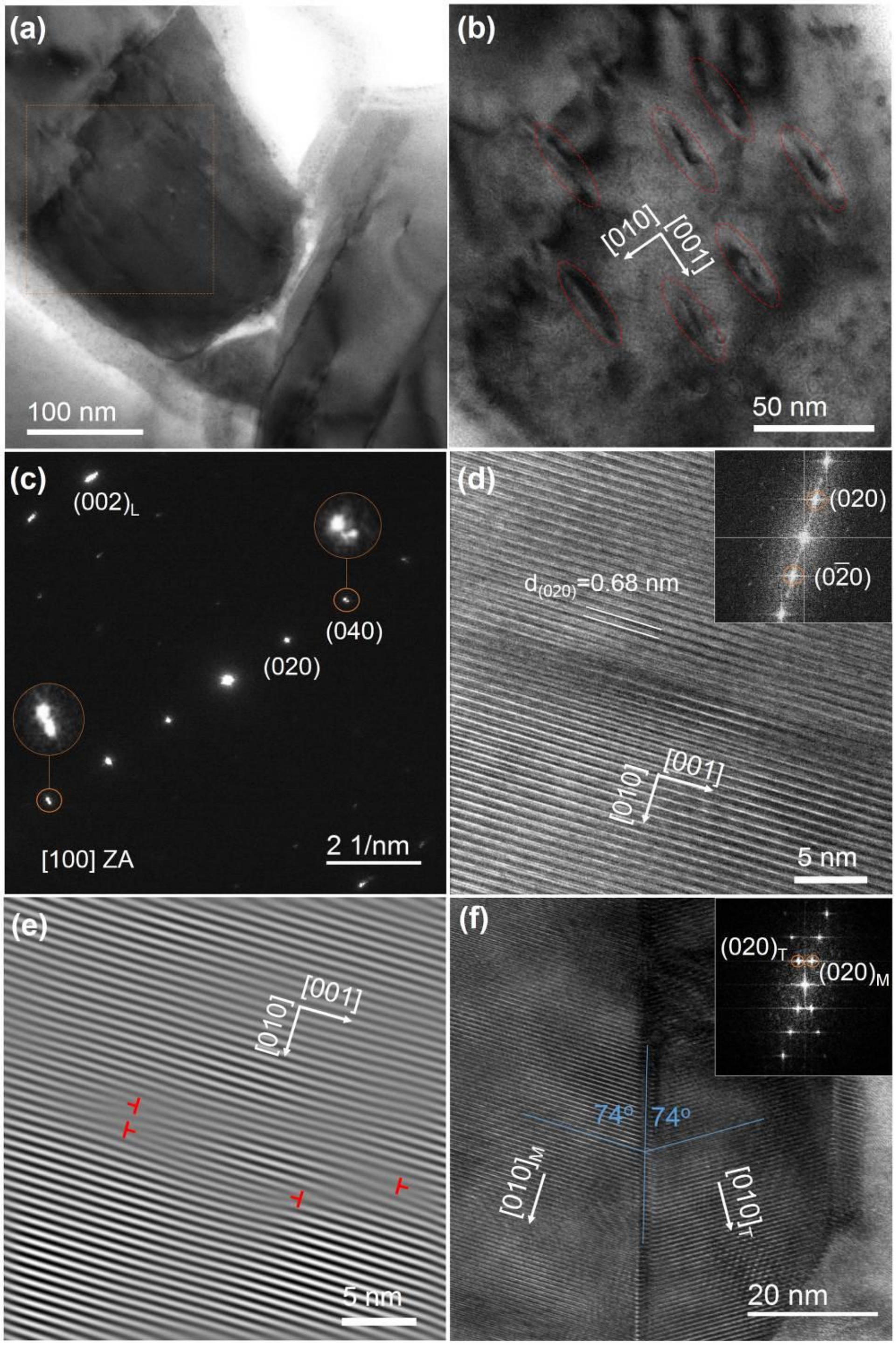

**Figure 4.** (Color online) (a) TEM image of the sintered $Sr_{14}Cu_{24}O_{41}$ sample 2 showing the nano-layered grains. (b) The corresponding bright-field TEM image of the square area in (a) under a two beam condition. (c) SAED pattern of the grain boundary in (a) obtained along the [100] zone axis (ZA). (d) HRTEM image taken between two parallel

nano-layers. The inset of (d) is its fast Fourier transformation (FFT) pattern. (e) The corresponding filtered inversed FFT of (d) from the (020) and (0-20) reflections. The red labels indicate the locations of edge dislocations. (f) HRTEM image taken from a twin boundary, which are labeled as M for matrix and T for twin. The inset shows the corresponding FFT.

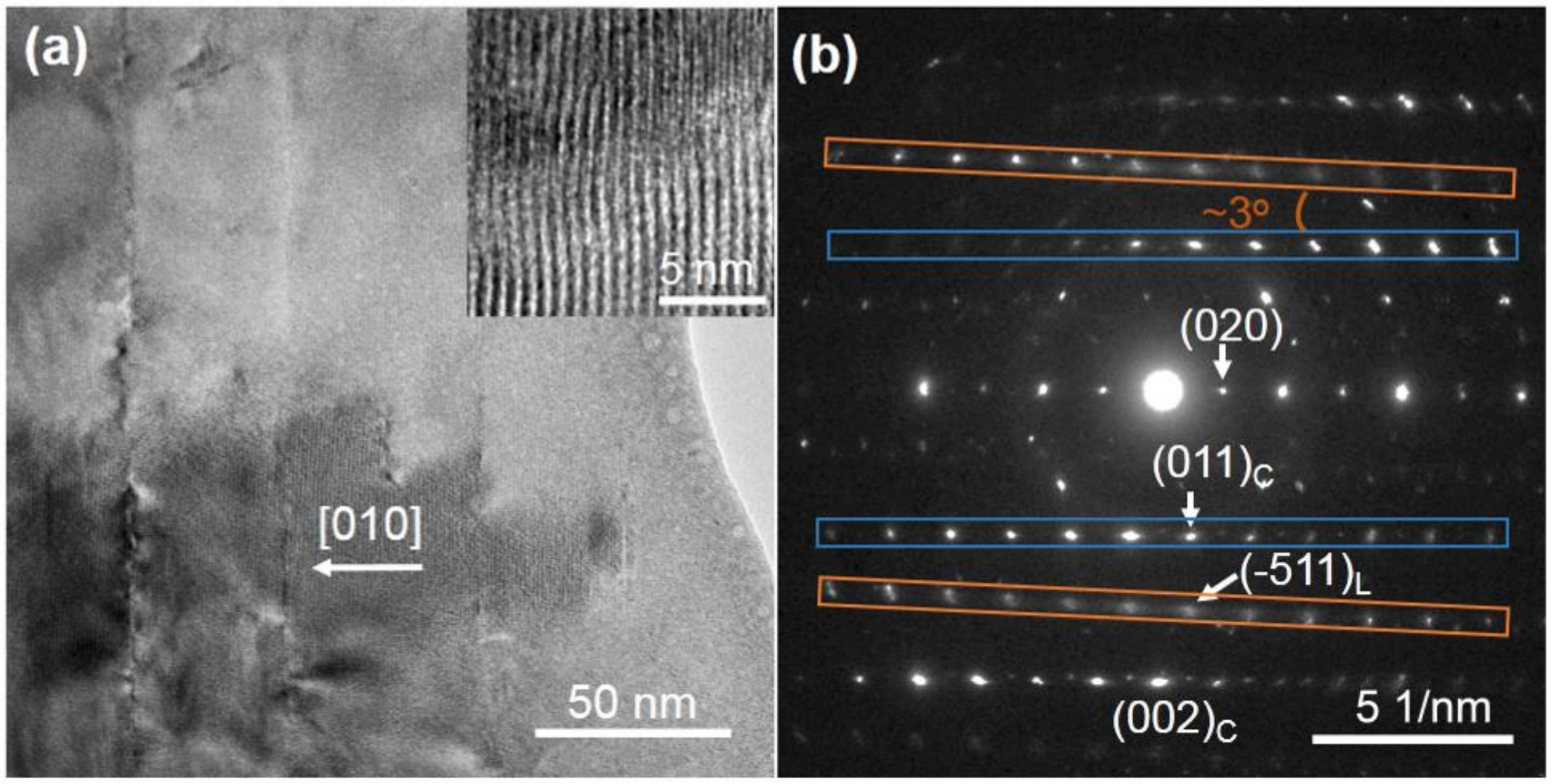

**Figure 5.** (Color online) (a) TEM images of the sintered $Sr_{14}Cu_{24}O_{41}$ sample 2 showing highly distorted nano-layers. The inset is the HTREM taken within a grain. (b) The corresponding SAED shows orientation anomaly between the chain and ladder sublattices.

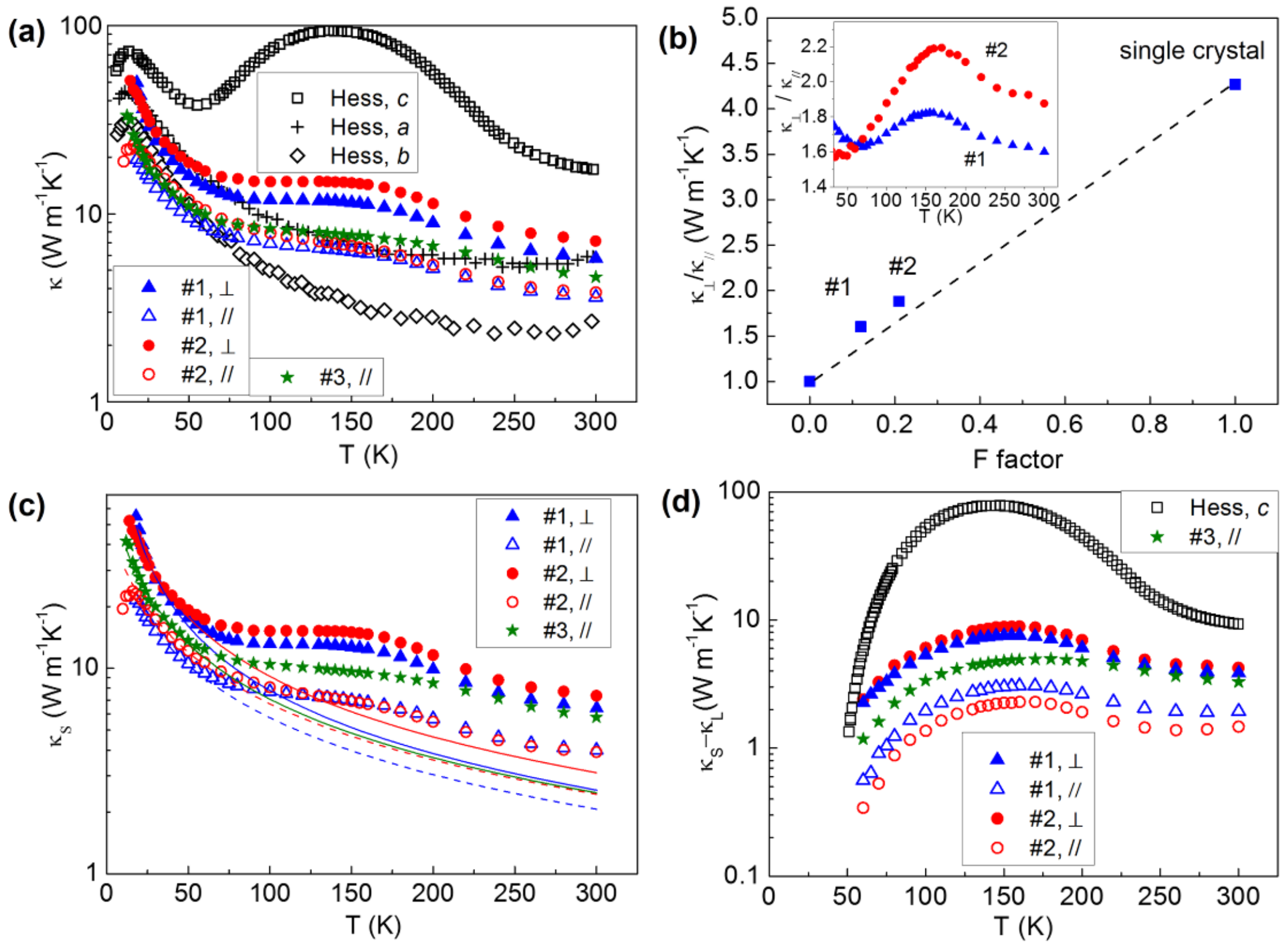

**Figure 6.** (Color online) (a) Thermal conductivity of two textured $Sr_{14}Cu_{24}O_{41}$ samples measured along two directions on logarithmic Y scale, in comparison with the thermal conductivity data of the single crystal reported by Hess *et al.*[10] and the polycrystalline sample with relatively random crystal orientation (sample 3) measured along the press direction. (b) The anisotropic thermal conductivity $\kappa_{\perp}/\kappa_{//}$ at 300 K as a function of the orientation factor. The dashed line is guide to the eye. The inset of (b) is the temperature dependence of $\kappa_{\perp}/\kappa_{//}$. (c) Solid thermal conductivity of the $Sr_{14}Cu_{24}O_{41}$ samples. The lines in (c) are the fitted lattice thermal conductivity data. (d) The extracted magnon thermal conductivity of the $Sr_{14}Cu_{24}O_{41}$ samples along two directions as a function of temperature. Shown for comparison is the reported magnon thermal conductivity of the single crystal along the *c* axis.[10]

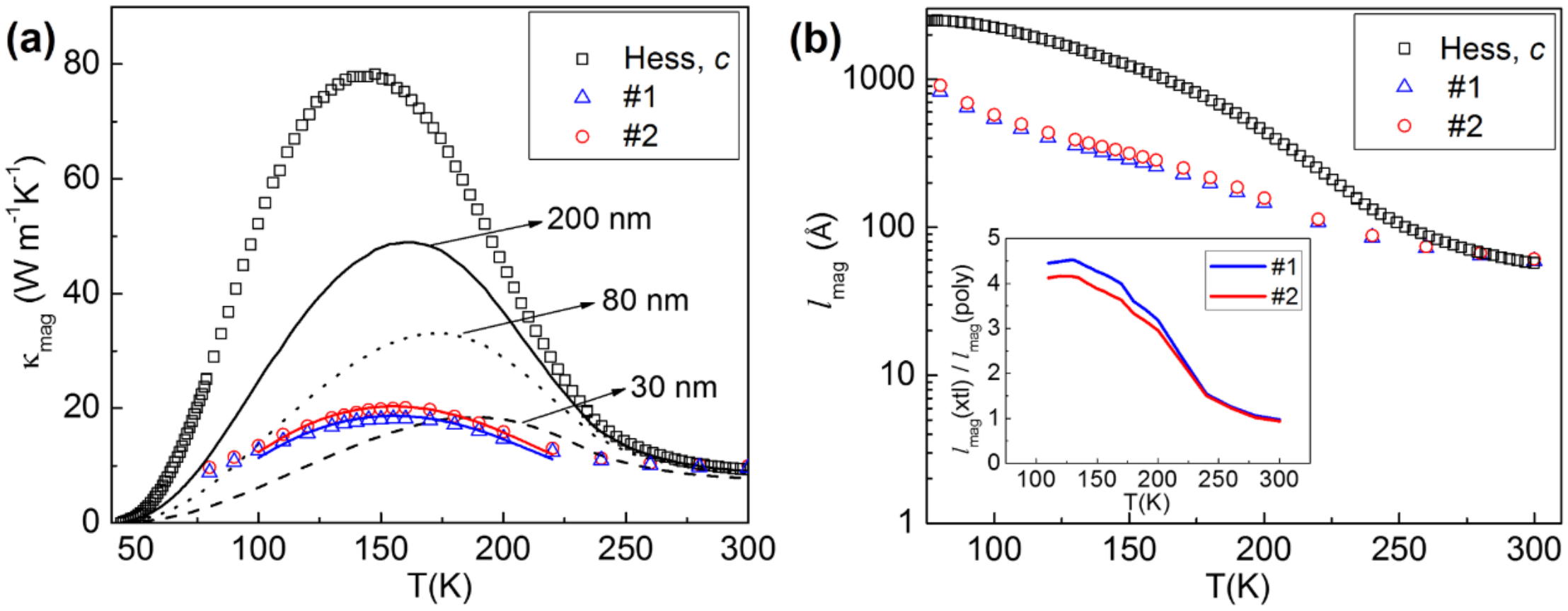

**Figure 7.** (Color online) (a) Intrinsic magnon thermal conductivity of the polycrystalline $Sr_{14}Cu_{24}O_{41}$ samples along the *c* axis. The black curves are the calculated magnon thermal conductivity for various boundary scattering MFPs without additional point defect scattering. The blue (for sample 1) and red (for sample 2) curves are the calculated magnon thermal conductivity with both boundary and defect scattering. (b) Calculated magnon MFPs of the polycrystalline $Sr_{14}Cu_{24}O_{41}$ samples as a function of temperature. Shown for comparison is the magnon MFP of the single crystal along the *c* axis.[10] The inset of (b) is the ratio between the magnon MFP of the single crystal and polycrystalline samples as a function of temperature.